# Progressive instability in circular masonry columns

M. Broseghini[a], P. Zanetti[b], A.D. Jefferson[b], M. Gei[b1]

[a] *Department of Civil, Environmental and Mechanical Engineering, University of Trento*
*Via Mesiano 77, 38123 Trento, Italy*
[b] *School of Engineering, Cardiff University, The Parade, Cardiff CF24 3AA, Wales, UK*

**Abstract**
The instability behaviour of eccentrically loaded circular masonry columns is investigated. Two approaches are considered for the analysis. One is based on a semi-analytical formulation of the relevant boundary-value problem for a no-tension material response; the other employs a plastic-damage-contact constitutive model, the CraftS model, to capture the complex microstructural behaviour of the material. The latter has been implemented in the finite element program LUSAS and has been already successfully employed to describe progressive instability in eccentrically loaded brickwork wallettes of rectangular cross section. Equilibrium paths and limit load estimations are computed for both analysis approaches for a range of column aspect ratios and load eccentricities. It is shown that the type of material response becomes less important for specimens with height-to-diameter aspect ratios greater than 7.5 and for loads applied to points in the kernel of the cross section, while for higher eccentricities the presence of a tensile strength increases considerably the limit load. The damage evolution predicted by the models is also investigated for selected cases, showing that the formulation based on the no-tension material is able to capture with good agreement the damaged zone of the column for loads with low eccentricities. For the same type of loading, a useful design formula is provided.

## 1. Introduction

The mechanical response of axially compressed brick masonry walls and columns is greatly affected by the eccentricity of the load. This dependency was well-known to ancient builders and has been investigated experimentally by a number of researchers [1-6]. As far as modelling is concerned, this problem can be tackled by assuming different hypotheses formulated to allow for the main features of masonry whose response is strongly nonlinear; these features being, for both bricks and mortar, a high resistance in compression, a low tensile strength and elastic-plastic damaging behaviour of the two materials. In addition, friction at their interface may be further considered. These features have been combined in different ways leading to various homogenization constitutive models for brickwork [7-11].

When a continuous beam-column model is required, the assumption that the material is elastic in compression and without any resistance in tension (the so called *no-tension material* model) can be postulated. In the past, many researchers have used these assumptions to estimate failure modes and loads of a range of compressed slender masonry structures. In particular, Sahlin [12], Yokel [13], and Frisch-Fay [14] were the first to investigate progressive instability, and maximum loads, in non-centred loaded pillars. Their approach was subsequently extended by the addition of self-weight effects (see Refs. [15,16]). The continuous assumption, which yields second-order differential governing equations to be integrated with the relevant boundary conditions, was then followed by the development of a 'discrete' approach [17] in which the pillar was divided into a number of finite blocks (or elements). The associated algebraic equations were formulated by imposing appropriate equilibrium conditions for the individual elements. The instability of homogeneous eccentrically loaded masonry pillars has been also investigated numerically, mainly using the Finite Element (FE) method. Ganduscio and Romano [18] developed a FE technique to consider no-tension materials with a non nonlinear response in compression. Brencich and Gambarotta [19] and Adam et al. [20] adapted a nonlinear plastic-damage-contact material model, originally developed for concrete, to successfully describe the behaviour of masonry wallettes.

---

[1] Corresponding author; email: geim@cardiff.ac.uk, tel: +44 29 2087 4316.

The majority of previous analyses considered prismatic specimens of rectangular cross section, although there have been a few previous investigations on the instability of circular cylinders [21,22]. In addition, Gurel [23] recently presented a study in which the load-bearing capacity of eccentrically loaded slender no-tension circular columns (including self-weight effects) was assessed with the discretised model proposed in [17]. In his paper, unbounded linear elastic compression was assumed and specimens with height-to-diameter ratios greater than 12.5 were considered. Moreover, results were compared with those obtained with a commercial FE package adopting the same constitutive model.

Based on the above introduction and state-of-the-art review, this paper describes four main contributions to the topic, as follows:

i) a method for determining the beam-column governing equations necessary to compute the compressive load-lateral displacement curve of a circular column comprising no-tension material loaded with a small or a large eccentricity, along with a numerical technique for solving these equations;

ii) the application of the above model to a set of circular columns loaded at various eccentricities, along with the computed equilibrium paths, limit loads, and the extent of the cracked zones to enhance the dataset provided in [23];

iii) the application of a fully coupled plastic-damage-contact (CraftS) constitutive model suitable for brickwork, implemented in a FE package, to the evaluation of the performance of the same specimens analysed in ii) so that an incisive comparison between different approaches can be made;

iv) a new formula which captures, with a good agreement, the limit load vs eccentricity behaviour of slender and relatively-slender columns, formed of a zero tensile strength material.

The paper is divided into six Sections. After the introduction, in Section 2, the constitutive models adopted in the analysis are introduced. In Section 3, the method to solve the analytical formulation of the beam-column boundary-value problems is illustrated, while the general setting for FE simulations is described in Section 4. Results and comparisons are contained in Section 5, followed by concluding remarks.

## 2. Models for the eccentrically compressed circular column

The features of the models considered in the paper are described in this Section. The prismatic column has diameter D and height L and x denotes the axial coordinate. The structure is assumed to be clamped at cross section x=0 and free at the top (x=L), where the resultant compressive load P is applied with the given eccentricity e (Fig. 1a).

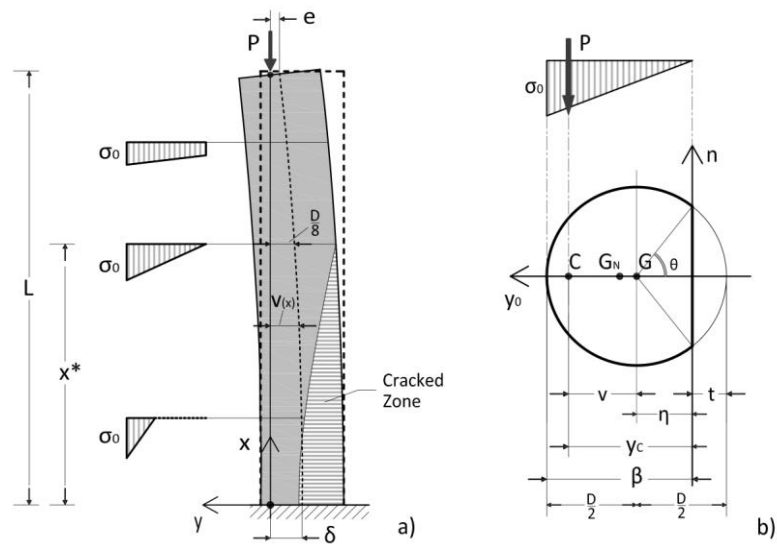

Fig. 1. Geometry and notation of the prismatic circular column: a) lateral view, where the quantities useful to describe the model based on no-tension material are also reported; b) cross section and stress distribution in the cracked zone in the case of no-tension material constitutive assumption (G is the centroid of the whole circle while $G_N$ denotes the centroid of the compressed part).

### 2.1 No-tension material with linear elastic compressive branch

A no-tension (NT) material is unable to withstand any tensile stress, while compression can be described in different ways, according to the nature of the solid. In our study, only vertical displacements at the base are constrained. Here, it is assumed that the uniaxial compression branch is unbounded linear elastic (E is Young's modulus). This hypothesis has led to closed-form solutions of the progressive instability of pillars with rectangular cross sections [12-14,24,25]. For a circular cross section, as shown later, an analytical solution is not available, but the resulting second order ODE can be readily numerically integrated. Although the NT model can be seen as the natural candidate for dry-joint masonry, its use for clay/mortar brickwork columns will be later assessed.

### 2.2 Plastic-Damage-Contact constitutive model (CraftS)

3D finite element simulations of the loaded column were carried out with the package LUSAS (ver 15.2) using a plastic-damage-contact constitutive model (named the CraftS model, originally formulated for concrete in [26]). An earlier version of this model was successfully employed to assess the behaviour of eccentrically loaded clay brickwork wallettes [20]. The model is able to simulate directional cracking, crack closure and the effects of frictional behaviour in compression, including triaxial confinement [27,28]. CraftS uses a crack-plane sub-model, which is based on a damage-contact formulation, and then incorporates the associated inelastic strains -derived from this sub-model- into a 3D elasto-plastic framework.

The triaxial plasticity component of the model employs a smooth triaxial yield surface due to Lubliner et al. [29], which is rounded using Willam and Warnke's [30] smoothing function. A work hardening hypothesis is used to simulate friction hardening behaviour in a manner similar to that used in the plasticity model presented in [31]. The friction hardening function is linked to a uniaxial compression curve for concrete such that the input parameters are the compressive strength ($f_c$) and strain at peak compression ($\varepsilon_c$). The friction hardening parameter (Z) has a range [$Z_0$,1], with $Z=Z_0$ defining the initial slope of the yield surface and $Z=1$ giving its maximum value. $Z_0$ defines the limit of the initial elastic region and is typically set to a value of 0.6, as explained in reference [26]. The other parameters used to govern the shape of the triaxial yield function are $f_c$ and the biaxial strength ratio ($b_r$) (see [26]). It noted that $b_r$ is defined as $f_b/f_c$, with $f_b$ being the maximum compressive stress in a biaxial test when the principal stresses are $\sigma_I=0$ and $\sigma_{II}=\sigma_{III}=-f_b$, assuming a tension positive convention.

The evolution of directional damage (or cracking) is governed by the following damage evolution equation, which was derived from the uniaxial stress-relative-displacement response of a notched axially loaded concrete specimen:

$$f_s(\zeta) = f_t \quad \forall \zeta \leq \varepsilon_K, \qquad f_s(\zeta) = f_t\, e^{-c_1\left(\frac{\zeta-\varepsilon_k}{\varepsilon_m-\varepsilon_k}\right)} \quad \forall \zeta > \varepsilon_K, \tag{1}$$

where, $f_t$ is the tensile strength, $\zeta$ is the effective strain parameter related to particular crack at a given element integration point and $\varepsilon_k$ is the strain at the peak of the uniaxial stress-strain curve, typically taken as 1.35 $f_t/E$, with the value of 1.35 corresponding to the parameter $a_k$ in reference [27]. Quantity $\varepsilon_m$ is the strain at the effective end of the uniaxial softening curve, which is computed from the element characteristic length and the specific fracture energy parameter ($G_f$) using the standard method of Bazant and Oh [32]; $c_1$ is a constant that takes the value 5, which was chosen to provide a close match with a range of uniaxial tension test data [26].

A particular feature of the CraftS model is the way that it deals with crack closure between rough surfaces [28]. Such behaviour is important for concrete in which aggregate interlock is an important loading carrying mechanism. The rough surface contact formulation considers the stress transferred across an open crack. The average slope of the crack surface asperities is governed by the parameter ($m_g$), with this slope being defined relative to a crack-plane normal. The effective height of the asperities, that controls the opening displacement beyond which shear can no longer be transferred across a crack, is set by the material parameter $m_{ful}$, which is applied as multiplier to the limiting opening displacement $u_{max}$. The quantity $u_{max}$ is the crack-opening displacement beyond which no significant direct tensile stress is transferred across the crack. The cracks that occur in masonry tend to be far smoother than those in concrete and this reflected in the rough surface contact parameters used for the present simulations. Mesh objectivity is achieved via the Bazant-Oh fracture-energy-based crack-band approach [32], as explained above.

The CraftS model is adopted for two types of simulation:
a) the column is considered as a *homogeneous* body with mechanical properties corresponding to two different materials (Table 1): one has properties similar to those of clay brickwork [20] (for which the simulations are indicated by FEC), the other resembles that of sandstone brickwork [33] (for which the simulations are indicated by FES). For simplicity, the tensile and compressive strengths for the latter material have been selected as four times those for clay brickwork;

b) the column is discretized into bricks and mortar layers (the materials are given in Table 1 in which the relevant brick properties are those presented in the first line) with the assumption that there are perfect interface transmission conditions between the two materials (corresponding simulations denoted by FEBM).

It is noted that the homogenized properties of clay masonry from (a) were used for the clay units in the discrete idealization (b). This is considered reasonable because, (i) the elastic properties of mortar and clay masonry are similar and therefore any minor variation in these values has little influence on the solutions and, (ii) the strength properties chosen for mortar were significantly weaker than those used for the homogenized masonry and therefore using higher strength values for the clay units in the discrete case would not have changed the outcome. The results from these simulations are presented in Section 5 of this paper.

|  | E [MPa] | v [-] | $f_c$ [MPa] | $f_t$ [MPa] | $G_f$ [N/mm] | $\varepsilon_c$ [-] | $b_r$ [-] | $m_g$ [-] | $m_{ful}$ [-] |
|---|---|---|---|---|---|---|---|---|---|
| 'Homogeneous clay brickwork' | 2000 | 0.15 | 13.8 | 3.3 | 0.1 | 0.008 | 1.15 | 0.4 | 3 |
| 'Homogeneous sandstone brickwork' | 2000 | 0.15 | 55.2 | 13.2 | 0.1 | 0.032 | 1.15 | 0.4 | 3 |
| Mortar | 1700 | 0.2 | 9.2 | 2.4 | 0.1 | 0.01 | 1.15 | 0.4 | 5 |

Table 1. Mechanical properties of material models used in FE modelling.

## 3. Analysis based on the no-tension material assumption

Depending on the eccentricity, different cases arise during loading: i) the column is compressed everywhere; ii) each cross section is partially damaged; iii) the column is fully compressed at the top and partially damaged below the unknown coordinate x* (this case is depicted in Fig. 1a). Field equations and boundary conditions for this analysis are described below.

### 3.1 Fully compressed column

When all cross sections are fully compressed and there is no damage, the displacement function $v(x)^2$ of the longitudinal axis (Fig. 1) is described by the equation

$$\frac{d^2v}{dx^2} + \frac{P}{EJ}v = 0, \qquad (2)$$

where P is the axial load and J is the second moment of area, namely, $J = \pi D^4/64$. Eq. (2) can be made dimensionless through the introduction of the normalized coordinate $\xi = x/D$ and the following non-dimensional quantities

$$\bar{v}(\xi) = \frac{v(\xi D)}{D}, \qquad \bar{P} = \frac{P}{ED^2}, \qquad \bar{J} = \frac{J}{D^4}. \qquad (3)$$

---
[2] The displacement w(x) of the longitudinal axis corresponds to $\delta$-v(x).

Note that $0 \leq \xi \leq \bar{L} = L/D$, while the function $\bar{v}(\xi)$ satisfies the inequalities $\bar{e} \leq \bar{v}(\xi) \leq \bar{\delta}$, where, similarly to other lengths involved in the problem, $\bar{e} = e/D$ and $\bar{\delta} = \delta/D$. Therefore, the dimensionless counterpart of (2) is

$$\frac{d^2\bar{v}}{d\xi^2} + \frac{\bar{P}}{\bar{J}}\bar{v} = 0. \tag{4}$$

To calculate the load-displacement relationship $\bar{P}(\bar{\delta})$, the following boundary conditions and compatibility condition must be employed

$$\frac{d\bar{v}(0)}{d\xi} = 0, \quad \bar{v}(\bar{L}) = \bar{e}, \tag{5}$$

$$\bar{v}(0) = \bar{\delta}. \tag{6}$$

### *3.2 Governing equation for a damaged pillar*

To describe the curvature of the axis when the structure is partially cracked, the following model is introduced. This is based on the deformation analysis of a small part of the column [13] (Fig. 1b)

$$\frac{d^2v}{dx^2} = -\frac{\varepsilon}{\beta} = -\frac{\sigma_0}{E}\frac{1}{\beta}, \tag{7}$$

where $\beta$ is the length of the compressed part of the cross section at x and $\sigma_0$ is the maximum stress (positive in compression) at the generic cross section while $\varepsilon$ is the corresponding longitudinal strain.

For a no-tension material beam, it is well known that $\sigma_0$ and $\beta$ are connected by the relationship [12]

$$\sigma_0(\theta(v)) = \frac{P\,\beta(\theta(v))}{S_n(\theta(v))}, \tag{8}$$

where the angle $\theta$ ($\theta \in [0, \pi[$) determines the position of the neutral axis and $S_n$ is the first order moment of the active part of the cross section with respect to axis n. Therefore, the governing equation becomes

$$\frac{d^2v}{dx^2} + \frac{P}{E\,S_n(\theta(v))} = 0. \tag{9}$$

To solve (9), $S_n$ must be expressed as a function of v which is accomplished via the function $\theta(v)$. To this end, the map $v(\theta) = y_C(\theta) - \eta(\theta)$, where $\eta$ is the distance between neutral axis and centroid G (see Fig. 1), is required. It is worth recalling that, with respect to the neutral axis, the coordinate of the pressure resultant point C is given by

$$y_C(\theta) = \frac{J_n(\theta)}{S_n(\theta)} \tag{10}$$

where, similarly to $S_n$, $J_n$ is the second-order moment of area of the uncracked portion of the cross section calculated with respect to n. $J_n(\theta)$ and $S_n(\theta)$ can be obtained by taking the difference between the same variables for the whole circle, respectively $J_{totn}(\theta)$ and $S_{totn}(\theta)$, and those associated with the complementary circular segment to the uncracked cross [$J_{csn}(\theta)$ and $S_{csn}(\theta)$], as follows

$$J_n(\theta) = J_{totn}(\theta) - J_{csn}(\theta) = \frac{1}{4}\pi D^4 \left(\frac{1}{16} + \frac{1}{4}\cos^2(\theta)\right) + \frac{D^4 \sin^6\theta}{18(2\theta - \sin 2\theta)} - \frac{1}{8} D^4$$
$$\cdot \left(\frac{2\sin^3\theta}{3(2\theta - \sin 2\theta)} - \frac{1}{2}\cos(\theta)\right)^2 (2\theta - \sin 2\theta) - \frac{1}{256} D^4 (4\theta - \sin 4\theta), \tag{11}$$

$$S_n(\theta) = S_{totn}(\theta) - S_{csn}(\theta) = \frac{1}{8}\pi D^3 \cos\theta + \frac{1}{8} D^3 \left(\frac{2\sin^3\theta}{3(2\theta - \sin 2\theta)} - \frac{1}{2}\cos\theta\right)(2\theta - \sin 2\theta), \tag{12}$$

so that $y_C(\theta)$ takes the form

$$y_C(\theta) = \frac{D(36\pi - 36\theta + 24(\pi - \theta)\cos 2\theta + 28\sin 2\theta + \sin 4\theta)}{8(12(\pi - \theta)\cos\theta + 9\sin\theta + \sin 3\theta)}. \tag{13}$$

As a final step, the required function is given by

$$v(\theta) = y_C(\theta) - \eta(\theta) = \frac{D(36\pi - 36\theta + 24(\pi - \theta)\cos 2\theta + 28\sin 2\theta + \sin 4\theta)}{8(12(\pi - \theta)\cos\theta + 9\sin\theta + \sin 3\theta)} - \frac{D}{2}\cos\theta. \tag{14}$$

As eq. (14) is transcendental, its analytical inversion is not known in closed form, therefore we introduce, through Fourier partial sums (two terms are considered and judged to be sufficient), the new function $v_a(\theta)$, which approximates $v(\theta)$ with a maximum error of about 2%,

$$v_a(\theta) = \frac{D}{64}\left[10 + 3\pi - 12\cos\theta - (3\pi - 10)\cos 2\theta\right]. \tag{15}$$

The inversion of $v_a(\theta)$ provides the required map $\theta(v)$ that is valid for $v \in [D/8, D/2]$, namely

$$\theta(v) = \arccos\left[\frac{-6 + \sqrt{36 - 120\pi + 36\pi^2 - 128\frac{v}{D}(3\pi - 10)}}{6\pi - 20}\right]. \tag{16}$$

The function $S_n(v)$, used in eq. (9), can be obtained by substituting (16) in (12) to yield the governing equation of the column in the cracked domain.

This equation can be made dimensionless through the use of the definitions given in list (3), to which we add $\overline{S_n}(\overline{v}) = S_n(\overline{v} D)/D^3$, to achieve the final expression

$$\frac{d^2\overline{v}}{d\xi^2} + \overline{P}\frac{1}{\overline{S_n}(\overline{v})} = 0. \tag{17}$$

It is worth noting that in (16) the displacement function $v(x)$ appears in dimensionless form since it is already divided by D.

For a large-eccentricity load ($\bar{e} > 1/8$), for which there is some damage along the entire length of the column, it is sufficient to integrate eq. (17) over the whole domain $0 \leq \xi \leq \bar{L}$, with boundary conditions (5) and the compatibility equation (6).

In the case where the column is partially cracked, the governing equations are

$$\frac{d^2\bar{v}}{d\xi^2} + \bar{P}\,\frac{1}{\overline{S_n}(\bar{v})} = 0 \quad (0 \leq \xi \leq \xi^*), \qquad \frac{d^2\bar{v}}{d\xi^2} + \bar{P}\,\frac{\bar{v}}{\bar{J}} = 0 \quad (\xi^* \leq \xi \leq \bar{L}), \tag{18}$$

which must be integrated in the relevant domains with the following boundary conditions

$$\frac{d\bar{v}(0)}{d\xi} = 0, \qquad \bar{v}(\bar{L}) = \bar{e}, \tag{19}$$

and interface continuity conditions at $\xi^*$ (i.e. at the interface between the cracked and the uncracked parts)

$$\bar{v}(\xi^*_-) = \bar{v}(\xi^*_+) = \frac{1}{8}, \qquad \frac{d\bar{v}(\xi^*_-)}{d\xi} = \frac{d\bar{v}(\xi^*_+)}{d\xi}. \tag{20}$$

The load-displacement equation $\bar{P}(\bar{\delta})$ may be found by imposing the compatibility condition at the foundation cross section, namely $\bar{v}(0) = \bar{\delta}$.

In all the cases analysed, the equations were integrated numerically with a tolerance of $\bar{e}/100$ in the dimensionless function $\bar{v}(\xi)$, obtaining the equilibrium path for each selected eccentricity $\bar{e}$.

## 4. Analysis based on finite element simulations with the CraftS model

Homogeneous clay and sandstone brickwork columns (sub-Section 2.2) were studied using three dimensional FE simulations with the program LUSAS (ver. 15.2). In all analyses, the load was applied through a steel plate, bonded on top of the column. The plate thickness-to-column height ratio was 1/50 and the steel Poisson's ratio was taken equal to that of the underlying material to avoid local effects due to strain mismatch. Coming back to the application of the load, the eccentric force is applied through an equivalent system of four forces aligned along a diameter of the upper surface of the plate: two opposite forces at the extremal nodal points of the diameter and two symmetric forces at the midpoints of the two radii aligned with the chosen diameter (see Fig. 2a). At the foundation cross section, the pair of points indicated in Fig. 2a were fully fixed, whilst for all other nodes the longitudinal displacements were set to zero. These conditions also ensured that the two in-plane shear stress components were also null.

In all cases, the columns were discretised using 8-noded isoparametric hexahedral elements (HX8). For the simulations with homogeneous materials (i.e. FEC and FES), D=100 mm. Two meshes were adopted (Fig. 2a), i) a coarse one, whose average size was approximately 11 mm x 11 mm x 16 mm, and ii) a finer one (average size, approximately 8 mm x 8 mm x 8 mm). The plots presented in Section 5 were computed with the coarse mesh, since a comparison between load-displacement paths obtained with the two discretizations for the case FES ($\bar{L} = 5$ and $\bar{e} = 0.07$) showed close agreement (within 1%) even at large displacements (Fig. 2b). The columns with bricks and mortar layers (FEBM, D =500 mm and L=2500.4 mm) were subdivided in 38 brick-mortar strata. Each brick (mortar layer) thickness measured 55 (10.8) mm. The average size of each element was approximately 8 mm x 8 mm x 7 mm.

In all simulations, an updated lagrangian formulation was adopted with an automatic step selection algorithm The tolerances on the L2 norms (i.e. the square root of the sum of the squares) of the residual force and iterative displacement vectors were 1% and 0.1% respectively, which resulted in the full solutions being obtained in 20 to 30 increments, with an average of 4 iterations per load increment.

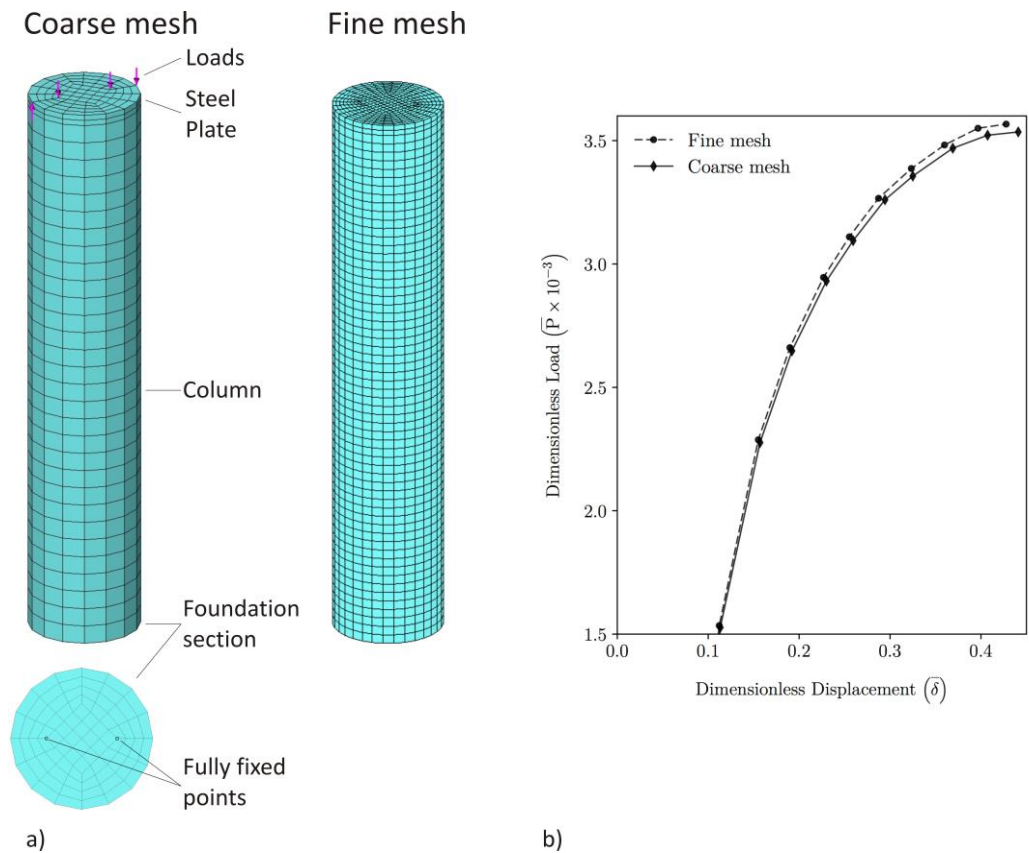

Fig. 2. a) Coarse and fine meshes for FEC and FES simulations, column with aspect ratio $\bar{L}$ = 5; b) convergence test for the FES simulation $\bar{L}$ = 5, $\bar{e}$ =0.07.

## 5. Results *and discussion*

### *5.1 Comparison between material models*

The results for columns with aspect ratios $\bar{L}$ = 5 and 7.5 are shown in Figs. 3a and 3b respectively for several values of the load eccentricity. The extent of the horizontal axis is limited to $\bar{\delta}$ = 0.5 as this corresponds to the limit condition for which the projection of the point of application of the load on the plane x=0 lies on the perimeter of the cross section (Fig. 1). In both plots, the lines associated with the dimensionless Eulerian buckling load $\bar{P}_{crE}=\pi^3/(256\bar{L}^2)$ and with the cracked/uncracked transition ($\bar{\delta}$ = 1/8) for NT simulations are shown for comparison. The equilibrium paths $\bar{P}(\bar{\delta})$ predicted by the NT model, based on the integration of eqs. (18), are shown with solid lines. For this model, part of the metastable branches beyond the peak loads are also reported. Interpolations of equilibrium points obtained from the FE simulations of homogeneous clay brickwork (FEC) columns are shown with long-dashed curves; those for homogeneous sandstone masonry (FES) with dash-dotted lines, whilst equilibrium paths for a column composed of bricks and mortar layers (FEBM) are denoted with dotted lines. In all cases studied with finite elements, the relevant curve terminates at the peak (this is not represented if it occurs at $\bar{\delta} > 0.5$), i.e. at the highest load reached by a successfully completed numerical simulation.

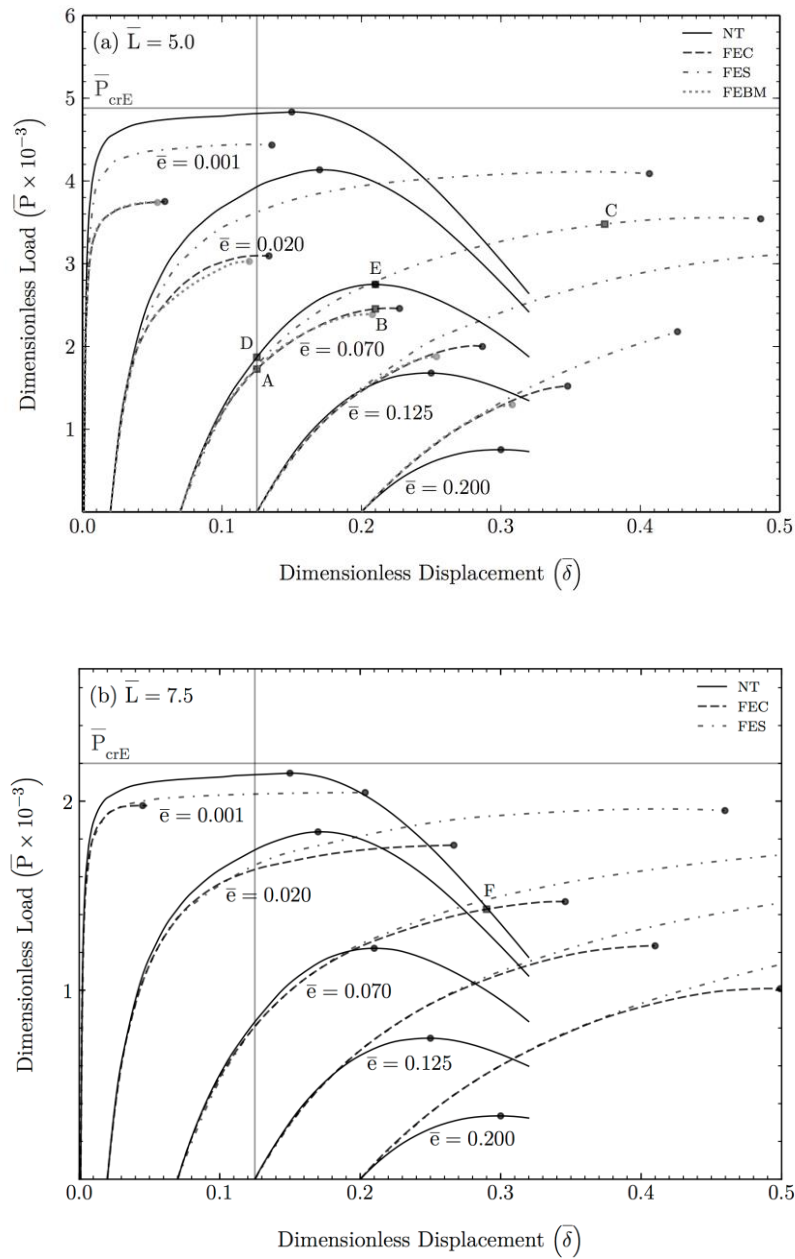

Fig. 3. Progressive instability curves of a circular column loaded eccentrically for different load eccentricities: a) $\bar{L}$ = 5; b) $\bar{L}$ = 7.5. NT: no-tension material results; FEC: FE homogeneous clay brickwork simulations; FES: FE homogeneous sandstone brickwork simulations; FEBM: FE simulations with brick and mortar discretised strata. Points A, B, C, D, E, F are explained in the text.

We first focus on the comparison between results pertaining to brick-mortar discretisation (FEBM) and those for homogeneous clay brickwork (FEC), as reported in Fig. 3a for $\bar{L}$ = 5. The loading paths are very close to each other for all the investigated loading cases. For the two higher eccentricities, the peak loads occur slightly earlier in the FEBM analyses due to local failures. However, it is observed that the homogeneous model captures very well the equilibrium path for the loaded column for all analysed eccentricities, therefore we can conclude that it is a sound model for this problem.

Turning attention to the comparison between the behaviour of clay brickwork columns and that of sandstone brickwork specimens, it is clear that the proximity of the two loading paths depends primarily on the slenderness of the structure and then upon the load eccentricity. For the least slender column, $\bar{L}$ = 5, and for the smallest $\bar{e}$ (=0.001), the two paths start to diverge just above $\bar{P}$ = 0.003 and the peak load for sandstone masonry is 22% higher than that of the clay brickwork column. Since both columns are only subject to compression, the difference can be explained with reference to a higher degree of plastic strain for the latter. Under these loading conditions (namely, $\bar{L}$ = 5 and $\bar{e}$=0.001), the NT curve can be seen as a limit envelope as the material always maintains the same stiffness, and plasticity is ruled out for this constitutive model. As an example of the behaviour of the CraftS model, along the loading history of the FEC simulation ($\bar{e}$ =0.001), the equivalent plastic strain (EPE in LUSAS FE package) starts to be non-zero approximately at $\bar{P}$ =0.0025 and is of the order of $2.2 \cdot 10^{-4}$ for $\bar{P}$ = 0.00338 and $8.6 \cdot 10^{-4}$ for $\bar{P}$ = 0.00373, almost at the peak load. For the corresponding FES computations, no plastic strains developed, showing that - for sandstone- brickwork instability is triggered by the structural behaviour associated with Euler buckling.

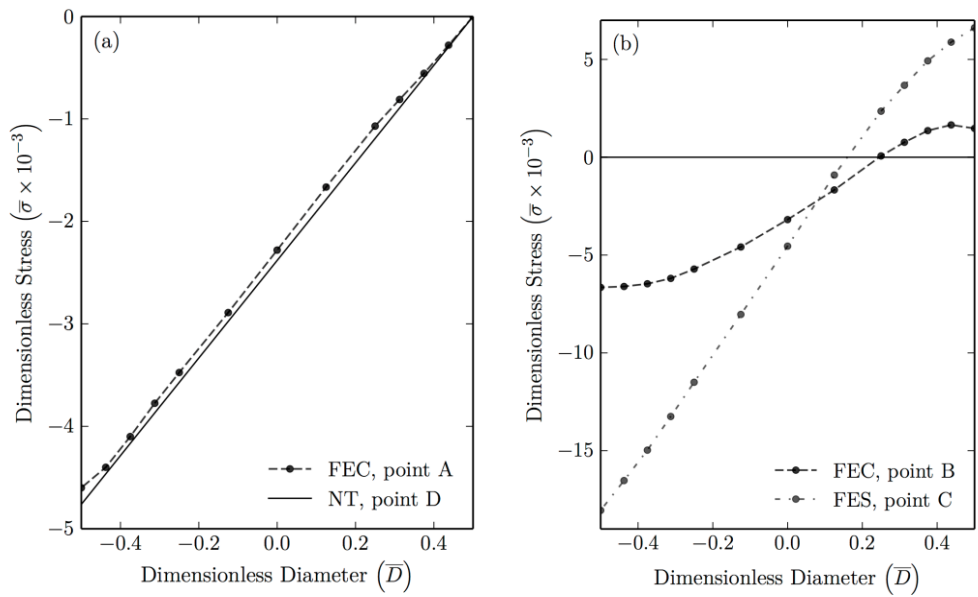

Fig. 4. Plots of the normalised normal stress distribution at the foundation cross section (x=0) obtained from FE simulations for (cf. points in Fig. 3a): a) FEC, point A (the linear plot for the NT material, point D - see also Fig. 6a- is also reported for comparison); b) FEC, point B, and FES, point C.

For higher eccentricities, the two paths are almost coincident up to 65-70% of the FEC peak load. However, sandstone masonry columns deform considerably more than the clay brickwork columns because the former have a tensile strength four times greater than the latter and thereby maintain equilibrium for longer, with respect to increasing δ. Fig. 4b (FES curve) shows the dimensionless normal stress ($\bar{\sigma} = \sigma/E$) at the foundation section (i.e. that at x=0) predicted by a FES simulation at point C in Fig. 3a, where it is manifest how wide is the area of the cross section subjected to tensile stress: the neutral axis is approximately at a distance equal to D/8 from the centre of the cross section.

For the more slender case, $\overline{L}$ = 7.5, the model predictions of two equilibrium paths at the same eccentricity are in closer agreement, with the peak loads differing by no more than 19% (this occurs for $\overline{e}$ =0.125). It is clear that for this aspect ratio (and presumably for higher aspect ratios) progressive instability is governed by the 'structural' behaviour of the column, for which the peculiar features of the material model become less important compared to the stocky specimen case. To support this statement, the equivalent plastic strain maps for two clay brickwork columns sharing the same eccentricity ($\overline{e}$ =0.07), but differing in aspect ratio, are represented in Fig. 5. The configurations, corresponding to points B and F in Fig. 3, share the same mean curvature. The maximum plastic strain for both columns is at the foundation section: for point B its value is 0.00337, while it is 0.00069 for point F. It has also been observed from the numerical simulations that, for point B (point F), the plastic strain drops to approximately 1/10 (1/3) of the value at just 90% of the load analysed indicating that for the less slender specimen there is a strong dependency of the behaviour on the constitutive response in the vicinity of the collapse.

The examination of plots in Fig. 3 reveals that, with respect to those obtained with the CraftS model, the column modelled with the NT material performs better at low eccentricity and worse at high eccentricity. The reasons have been partially explained earlier: when the load is almost centred, the cross section is fully compressed, the NT material -in contrast to the behaviour of the CraftS nonlinear material model- maintains the same elastic modulus independently of the stress and the structure does not experience any stiffness reduction due to plasticity at high stresses. At high eccentricities, the absence of any resistance in tension significantly reduces the structural stiffness, such that the limit condition for equilibrium is reached at a lower load compared to that obtained with the nonlinear material model. Whilst the former effect is amplified for columns with low slenderness ratios, the latter is enhanced by an increase of the aspect ratio. We note, however, that for the more slender aspect ratio, $\overline{L}$ = 7.5, the peak load is only slightly overestimated by the NT material with respect to that computed with the CraftS model at low eccentricities ($\overline{e}$ =0.001, 0.02). At higher eccentricities, but still lying in the kernel of the circular cross section ($\overline{e}$=0.07, 0.125), the peak loads computed with the no-tension assumption are slightly lower than that predicted by CraftS, while they are markedly lower for very high eccentricities ($\overline{e}$ =0.2 in our case). These findings show that for slender specimens ($\overline{L} \geq 7.5$) the NT model can be used to estimate, with reasonable accuracy, the equilibrium paths corresponding to progressive instability and maximum forces for loads whose resultant is centred within the kernel. It is also noted that, similar to the previous results for columns of rectangular cross section (see [12,25]), the critical load for the no-tension material always occurs when the column is cracked, even for very small initial eccentricities.

It is interesting now to compare the deformed configurations of the column predicted by NT and homogeneous clay brickwork (FEC) material models at two relevant stages of the equilibrium path computed for $\overline{e}$ =0.07 and $\overline{L}$ = 5 (Fig. 3a): the first stage considered is when a crack first arises for the no-tension material structure ($\overline{\delta}$ = 1/8, points marked by letters D and A respectively for NT and FEC); the second stage is at the maximum load, again for NT (point labelled E) compared to point labelled B for the FEC simulation at the same dimensionless displacement $\overline{\delta}$. Fig. 6 shows, for the NT material model, the deformed configuration and normal stress distribution (in dimensionless form) at the foundation cross section (x=0) for both points D (part a) and E (part b); for the latter, the extent of the cracked zone is also indicated. The stress state in Fig. 6a can be compared with that of Fig. 4a which has been calculated for the FEC simulation (point A) at the same cross section. The maximum compressive stress of the latter is slightly lower than that of the former as the load applied to the column at the configuration corresponding to point A is 6% lower than that for point D.

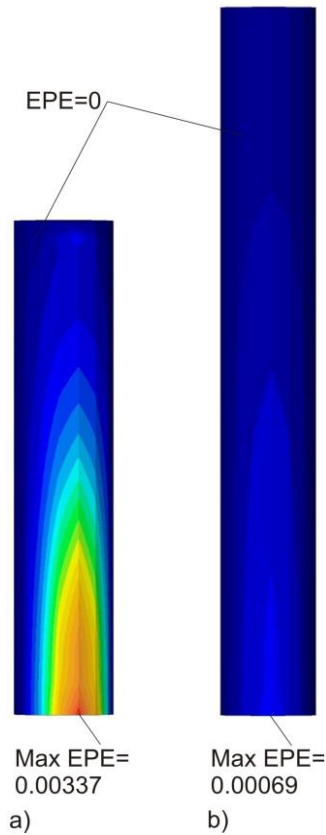

Fig. 5. Plots of the equivalent plastic strain (EPE in LUSAS FE package) for FEC simulations for (cf. points in Fig. 3): a) $\bar{L}$ = 5, $\bar{e}$ =0.07, point B; b) $\bar{L}$ = 7.5, $\bar{e}$ =0.07, point F. Note that the two columns share the same curvature. To show the plastic strain distribution in the column, in a) the range [0, 0.0034] has been divided in twenty intervals (0.17·10$^{-3}$ each) of different colours. In b) the same colour map is adopted.

A similar comparison can be performed for the columns whose configurations are represented by points E and B in Fig. 3a. While for the former the foundation cross section is partially cracked (Fig. 6b), so that only the compressive part contributes to the equilibrium of the column, the latter (Fig. 4b, FEC curve) shows a significant portion of the cross section subjected to tensile stresses, part of which displays softening. The resistant moment brought about by these tensile stresses explains why the maximum load for the clay brickwork occurs at higher displacements.

To conclude the analysis, we have computed, for point B, the extent of the volume of the column where the FEC simulation displays positive longitudinal normal stresses (see Fig. 6c) so that a comparison can be made with the data which describe, for point E, the cracked part of the NT model, namely $\xi^*$=3.76 and $\bar{t}$=0.264 (Fig. 6b). For the homogeneous clay brickwork model, the corresponding values of the parameters are $\xi^*$=3.97 and $\bar{t}$=0.233. This comparison simply shows that the NT model is able, to some extent, to predict the parts of a real brickwork column that would be subjected to compressive and tensile normal stresses where the latter are present only in a small volume of the specimen.

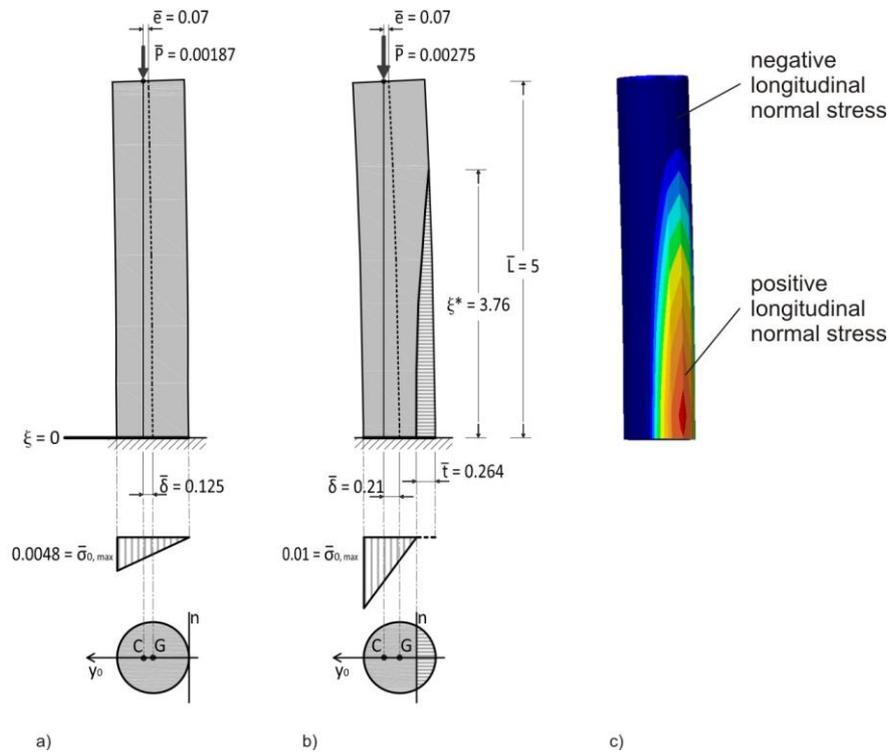

Fig. 6. Deformed configurations of the column with aspect ratio $\overline{L}$ = 5, $\overline{e}$ =0.07 for (cf. points in Fig. 3a): a) NT, point D; b) NT, point E; c) FEC, point B, dark blue: volume where the longitudinal stress is negative; other colours: volume where the longitudinal stress is positive; the stress distribution at the foundation cross section for this case is represented in Fig. 4b.

### *5.2 Critical load vs eccentricity*

An effective way to compare the load-bearing capacity of the examined columns is to summarise the data in a graph where the dimensionless critical load ($\overline{P}_{cr}$) is plotted against the parameter $\overline{e}$. For the NT material, this is provided in Fig. 7, where the maximum loads (made dimensionless though division by $\overline{P}_{crE}$) of the load-displacement paths analysed in Fig. 3 are reported. In the same figure, the case $\overline{L}$ =10, not previously discussed in detail, is also reported. Note that in this diagram, the three points for the three aspect ratios at the same eccentricity almost coincide. In Fig. 7, some numerical data extracted from Fig. 17 of Ref. [23], which refer to $\overline{L}$ =15, are also reported. In that paper, the self-weight is always considered; however, here these contributions have been estimated, adopting as a guideline the method proposed in [17], and eliminated.

The data processing has revealed that the critical load-eccentricity trend follows, with good agreement, the empirical law

$$\overline{P}_{cr} = \overline{P}_{crE} \exp(-\alpha\, \overline{e}), \tag{21}$$

where $\alpha$ = 8.79. A similar plot for the prismatic pillar with rectangular cross section was presented by De Falco and Lucchesi [25], where eq. (21) can be applied for moderate eccentricities with $\alpha \cong 7$. Though eq. (21) is not an exact result, it constitutes a useful tool for design purposes highlighting an interesting unifying trend of the peak load taking Euler buckling as a reference. In the range of eccentricities investigated, we expect eq. (21) to be valid for $\overline{L} > 5$, while less indicative for $\overline{L} < 5$, for which the progressive instability is strongly influenced by the material behaviour. As a final remark, it is interesting to observe that while the condition $\overline{P}_{cr} = \overline{P}_{crE}/2$ is reached, for the circular column, for $\overline{e} \approx 0.08$, the same condition is attained for a pillar of rectangular cross section at $\overline{e} \approx 0.10$.

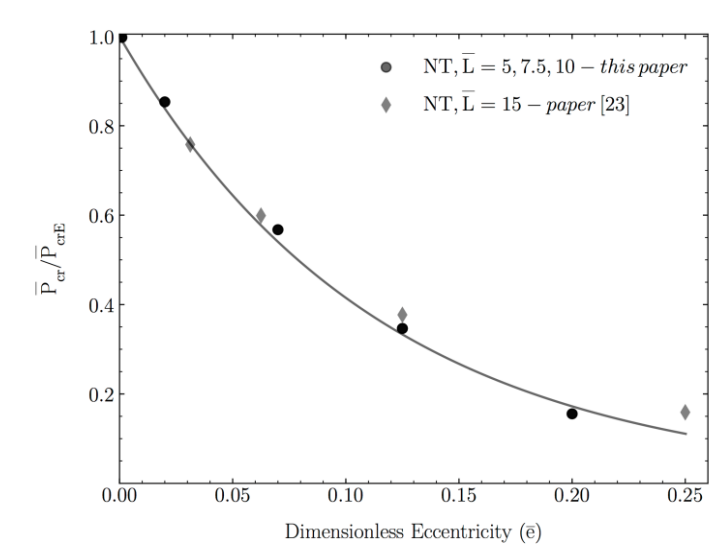

Fig. 7. Progressive instability of a circular column loaded eccentrically: dimensionless critical load vs. dimensionless initial eccentricity. The solid line represents eq. (21).

## 6. Conclusions

In this paper, we have analysed several aspects of the instability of homogeneous masonry circular columns loaded with a compressive longitudinal eccentric force. Our goal was to explore the prediction of maximum loads along load-lateral displacement equilibrium paths for constitutive laws which represent the main features of brickwork. In particular, no-tension material (with unbounded linear elastic behaviour in compression) and CraftS models were considered, the latter implemented in a FE package. A numerical technique for solving the nonlinear governing equations of a beam-column whose material obeys the former model has been proposed. The method can be extended to a generic cross-section shape with possible dependency on the longitudinal axis.

Several load eccentricities for columns of two height-to-diameter aspect ratios $\overline{L}$ have been analysed in detail. For $\overline{L}=5$, representing a relatively slender specimen, the maximum forces for low-eccentricity loads (i.e. the resultant lies in the kernel of the cross section) are strongly influenced by the elastic-plastic behaviour of the material. We have shown that the homogeneous column obeying the CraftS model may well replace that made of distinct brick and mortar layers. Moreover, the no-tension material model may significantly overestimate the limit load for clay brickwork columns, while it gives more realistic predictions for high-strength materials such as sandstone masonry. For more slender pillars ($\overline{L} \geq 7.5$), the structural behaviour dominates the progressive instability so that nonlinear material behaviour in compression becomes less important. For this class of masonry column, and again for low eccentricities -that constitute the most relevant conditions for this kind of structure in the engineering practice- the assumption of zero tensile strength leads to good estimates of the peak loads. However, this assumption becomes inaccurate for high eccentricities, due to its under-prediction of flexural capacities and, consequentially, maximum loads.

The damage evolution predicted by the two models is also investigated for selected cases, showing that the formulation based on the no-tension material is able to capture with good agreement the damaged zone of the column for loads with low eccentricities. In addition, for specimens following NT assumptions, a useful formula for design purposes is proposed, which relates the load-carrying capacity of the column to the Euler buckling load of the structure and is valid for aspect ratios larger than 5 and eccentricities lower than 25% of the column diameter. This outcome complements that proposed in Ref. [25] for pillars with rectangular cross sections modelled with the same type of material.

Though based on simplified assumptions, the no-tension material model may be particularly reliable for the computation of the collapse of compressed historic masonry columns where the poor quality of the mortar justifies its use.

**Acknowledgements**
The authors would like to thank the FE company LUSAS (www.LUSAS.com) for their cooperation in this work. Support from the EU FP7 project ERC-AdG-340561-INSTABILITIES is also gratefully acknowledged.